\documentclass[%
amsmath,amssymb,
final,
]{elsarticle}
\usepackage{graphicx}% Include figure files
\usepackage{dcolumn}% Align table columns on decimal point
\usepackage{bm}% bold math
\usepackage{color}
\usepackage[utf8]{inputenc}

\begin{document}

%%\title{Optical force on one-dimentional PT-symmetric structures}

\title{Dynamic behavior of mechanical cloaks designed by direct lattice transformation}
\author[label1,label2]{Muamer Kadic}
\ead{muamer.kadic@femto-st.fr}
\author[label2,label3]{Martin Wegener}
\author[label4]{Andr\'e  Nicolet}
\author[label4]{Fr\'ed\'eric Zolla}
\author[label4]{S\'ebastien Guenneau}
\author[label4]{Andr\'e Diatta}

\address[label1]{Institut FEMTO-ST, UMR 6174, CNRS, Universit\'{e} de Bourgogne Franche-Comt\'{e}, 25000 Besan\c{c}on, France}
\address[label2]{Institute of Nanotechnology, Karlsruhe Institute of Technology (KIT), 76128 Karlsruhe, Germany}
\address[label3]{Institute of Applied Physics, Karlsruhe Institute of Technology (KIT), 76128 Karlsruhe, Germany}
\address[label4]{Aix$-$Marseille Univ, CNRS, Centrale Marseille, Institut Fresnel, 13013 Marseille, France}

%\date{\today}

\begin{abstract}
Steering waves in elastic solids is more demanding than steering waves in electromagnetism or acoustics. As a result, designing material distributions  which are the counterpart of optical invisibility cloaks in elasticity poses a major challenge. Waves of all polarizations should be guided around an obstacle to emerge on the downstream side as though no obstacle were there. Recently, we have introduced the direct-lattice-transformation approach. This simple and explicit construction procedure led to extremely good cloaking results in the static case. Here, we transfer this approach to the dynamic case, i.e., to elastic waves or phonons. We demonstrate broadband reduction of scattering, with best suppressions exceeding a factor of five when using cubic coordinate transformations instead of linear ones. To reliably and quantitatively test these cloaks efficiency, we use an effective-medium approach.
\end{abstract}

%\pacs{45.90.+t, 46.40.-f}
%\keywords{cloaking, lattice metamaterials, Cosserat elaticity, phonons}

\maketitle	
\section{Introduction}
An invisibility cloak has long been considered as unachievable in practice. Over a decade ago, the concept of transformational \cite{Pendry2006} and conformal \cite{Leonhardt2006} optics were proposed \cite{Pendry2006,Leonhardt2006}. These can be connected to computational tools used in twisted fibres \cite{Nicolet2004b} and in electrical engineering \cite{Binns1963}. In recent years, different types of cloaks have been realized. This includes electromagnetic/optical invisibility cloaks \cite{Schurig2006,Zolla2007,Gabrielli2009,Renger2010,Ergin2010,Kadic2011,Kadic2012b,McCall2018}, cloaks for water waves \cite{Farhat2008,Berraquero2013,Dupont2015,Zareei2015,Dupont2016} and airborne sound \cite{Popa2011,Zhang2011,Sanchis2013}, vibration cloaks \cite{Brun2009,Farhat2009,Amirkhizi2010,Stenger2012,Misseroni2016}, static electric cloaks \cite{Yang2012}, and thermal cloaks \cite{Guenneau2012,Schittny2013}. However, cloaking elastic waves in solids is far more challenging as the Navier equations are generally not form invariant under general coordinate transformations \cite{Milton2006,Brun2009,Norris2011}, or require some special treatment to make them covariant \cite{Steigmann2007,Yavari2008}. Nonetheless, two large-scale experiments have shown that extreme control \cite{Brule2017} and protection \cite{Brule2014} can be achieved for surface Rayleigh waves. Rayleigh waves, which are polarized normal to the surface, can be converted into in-plane bulk shear waves by a metasurface \cite{Colombi2017,Palermo2018}. This finding suggests that achieving cloaking of in-plane elastic waves might be useful to control Rayleigh waves with potential applications ranging from ultrasonic sensing to earthquake protection in civil engineering.

\begin{figure}[h!]
	\begin{center}
\includegraphics[width=8.5cm,angle=0]{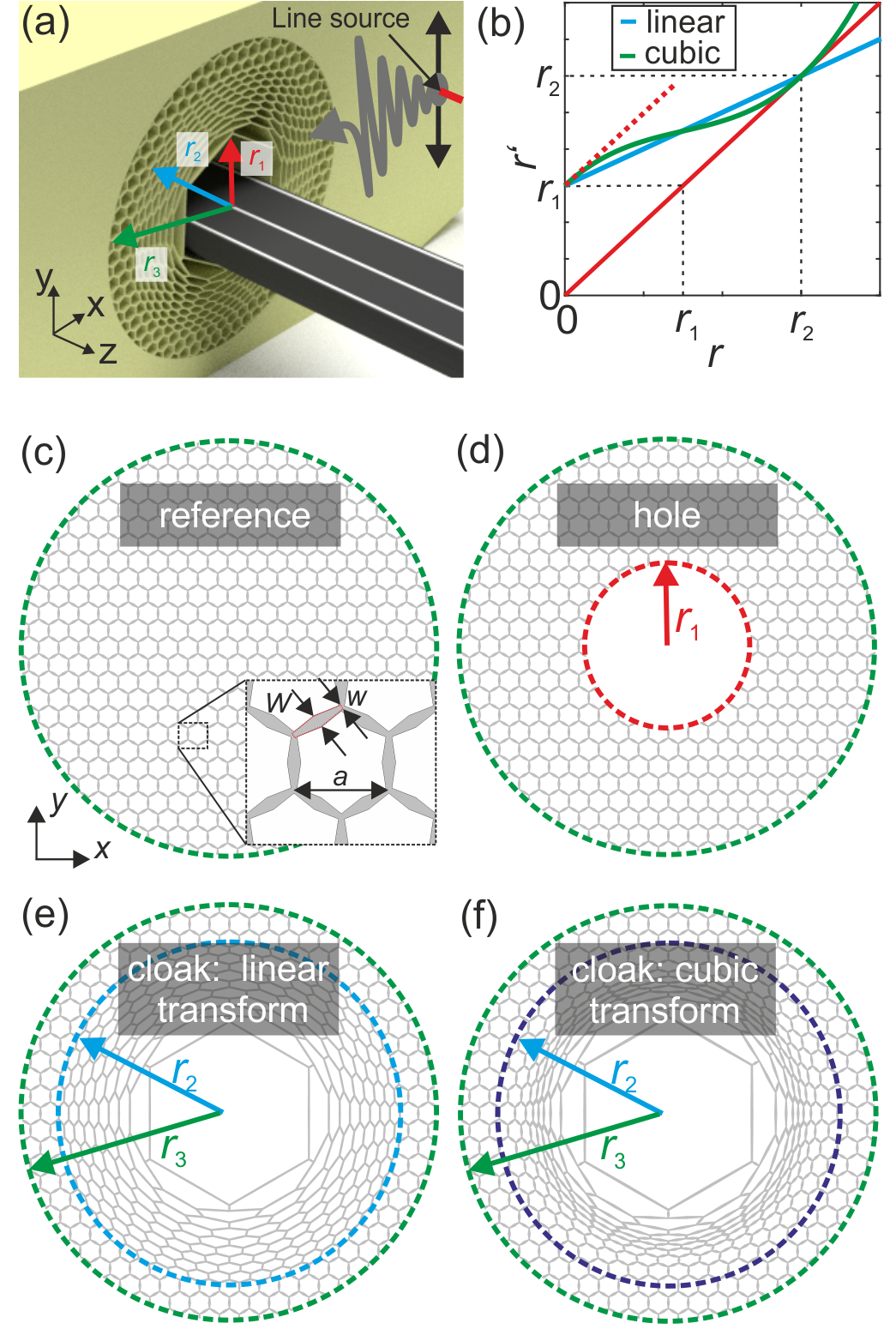}
\caption{(a) Principle of a lattice in-plane-shear cloak. An incident wave generated by a line source (in red) is guided around the region to be hidden. The structure is invariant along the $z$-axis. We start with the reference medium shown in (c). We cut a stress-free hole inside it (shown in (d)). This structure will be refered to as the obstacle. We design two different elastic cloaks by using the direct-lattice-transformation approach. These  two follow the 2 geometrical transformations depicted in panel (b). The red dashed line shows the derivative of the cubic transformation at $r=0$. The inner radius of the cloak is $r_1$, its outer radius $r_2$, and $r_3$ is the radius of the lattice medium. Outside of the disc of radius $r_3$, an effective elastic medium starts. $w$, $W$ and $a$ are the geometrical parameters of the lattice cell. For the homogeneous lattice medium, we choose $a=\sqrt{3}h$, $w/W=0.8$, $h/w=8$, and we fix $h=4 \; \rm cm$. For the cloak derived from a linear (resp. cubic) coordinate transformation, (e) (resp. (f)), we derive $w_0/w=0.7$ and $W_0/W=1.35$ ($w_0/w=0.81$ and $W_0/W=1$). For the cubic transformation, it is possible to choose the parameters such that the derivatives at $r_1$ and $r_2$ equal unity. This choice avoids an impedance mismatch at these interfaces.}
\label{Figure1}
\end{center}
\end{figure}

From the mathematical point of view, for elastic waves in solids (tensorial), the situation is much more complex than for acoustic waves in fluids (scalar). The underlying continuum-mechanics equations arising from Newton's law and Hooke's law are not form invariant under spatial transformations \cite{Milton2006,Brun2009,Norris2011}, at least not for the elasticity tensors of ordinary materials -- in sharp contrast to all examples listed in the introduction. Intuitively, waves in elastic solids usually have two transverse and one longitudinal polarizations, whereas electromagnetic waves are usually only transverse and acoustic waves are only longitudinal. Plainly speaking, it is difficult to simultaneously guide all three elastic wave polarizations around an obstacle as though no obstacle was there because the wave properties depend on the polarization. In fact, it generally requires so-called Cosserat materials to accomplish the goal of cloaking \cite{Brun2009,Norris2011}. While the corresponding mathematical description in terms of effective material parameters has been worked out for the 2D \cite{Brun2009} setting and later for the 3D case \cite{Diatta2014}, it is presently not clear how the resulting spatial rank-four tensor distributions can be approximated by specific practical microstructures. Likewise, it is not clear either how well these microstructures would eventually work. 
%\vspace{-0.5cm}
\begin{figure}[h!]
	\begin{center}
	\includegraphics[width=10cm,angle=0]{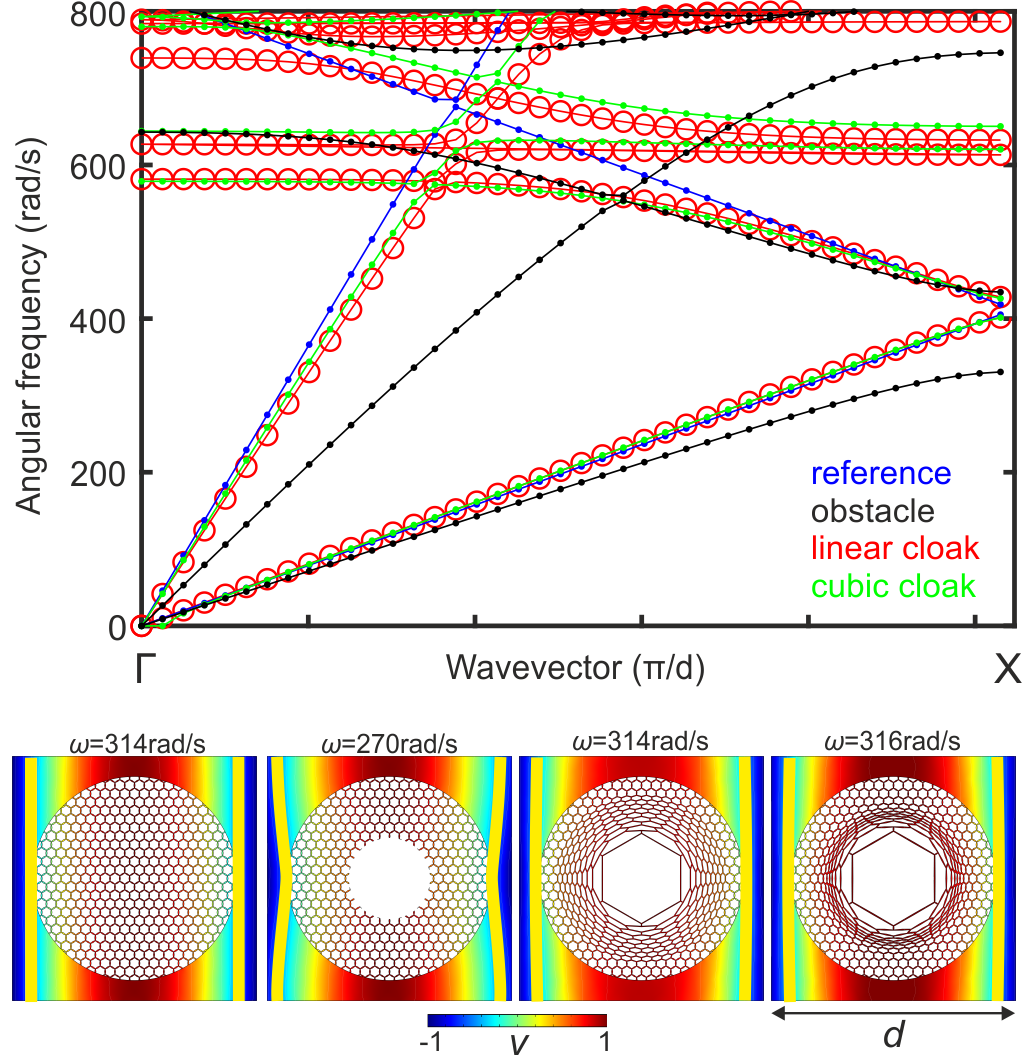}
	\caption{Dispersion diagram for the supercells containing the lattice medium and the surrounding effective medium. Four cases are depicted: Reference in blue, obstacle in black, the optimized linear cloak (red), and the optimized cubic cloak (green). The insets at the bottom correspond to in-plane shear eigenmodes for a supercell lattice composed of the reference, the obstacle, and the lattice cloak, respectively, arranged into a periodic super lattice. We fix the wavevector $k=2/d$ where $d=25a$. A strong deviation of the eigenfrequency is observed for the obstacle (270\,rad/s), whereas both cloaks display nearly the same eigenfrequency as the reference sample. Here, $\rm v$ is the y-component of the displacement field.}
	\label{Figure2}
\end{center}
\end{figure}

\section{Lattice cloak design}
Recently, we have introduced the direct-lattice-transformation design approach \cite{Buckmann2015}. This approach avoids speaking about effective material properties altogether. It rather starts from a micro-lattice of points connected by some sort of elastic elements. Next, the lattice points are subject to a wanted spatial coordinate transformation. As a result, the lengths of the connecting elements change. The basic idea of the approach is to adjust the width of each element such that its Hooke's spring constant -- corresponding to compression along the element axis –- stays constant. The approach ignored the mass density, which is justified because the mass density does not enter in the static case.  On this basis, excellent cloaking results have been obtained in the static case for a broad range of parameters. Such cloaks might find applications in stabilizing structures into which a hole needs to be punched at some point to allow for some feedthrough. The hole alone obviously weakens the support structure mechanically. The cloak guides the mechanical force field around the feedthrough such that the stabilizing structure has nearly the same properties as before.

Given this past success, the next question is whether the design approach can be expanded towards finite frequencies, i.e., to elastodynamic waves. It is clear that the behavior in the low-frequency limit must connect continuously to the static case. It is also clear that the cloak will fail once the frequency of the wave is so high that the corresponding wavelength becomes comparable to or even smaller than the lattice constant of the micro-lattice. In this limit, the approximation of the connections in terms of simple Hooke's springs breaks down because the connections are stretched or compressed inhomogeneously within themselves. It is less clear, however, how the distribution of masses affects the behavior. It is also not clear what the resonance eigenfrequencies of the cloak and the hole within it are and how these resonances would affect the behavior. 

Furthermore, in the static case, the impedance mismatch at the cloak's boundary is irrelevant. In contrast, in the dynamic case, wave reflections at this interface can lead to the dominant scattering contribution. We will specifically address this aspect by going from a linear coordinate transformation in polar coordinatess ($r$, $\theta$), defined by \cite{Pendry2006} $$r'=\frac{r_2-r_1}{r_2} \; r + r_1$$ and used in the static case, to a higher order transformation \cite{Cai2007} such as cubic one, defined by $$r'=a_3\, r^3 +a_2\, r^2 + a_1\, r +a_0$$ in the dynamic case. Here, we use the following parameters:
\begin{eqnarray} 
a_3=\frac{2 r_1}{r_2^3}, \; a_2=\frac{-3 r_1}{r_2^2}, \; a_1= 1, \; {\rm and} \; a_0=r_1. \nonumber
\end{eqnarray}
We note that if we differentiate the cubic transformation we get $$\frac{\partial r'}{\partial r}=3\, a_3 \, r^2+2\,a_2\, r +a_1$$ which is a necessary condition for an adiabatic transformation between the cloak and the surrounding: $$\frac{\partial r'}{\partial r}(0)=a_1=1=\frac{\partial r'}{\partial r}(r_2)$$.

In Figure \ref{Figure1}, we depict the general concept of the embedded lattice cloak inside of a homogeneous isotropic linear elastic material. The designed cloak is placed in the homogenized environment and an in-plane shear wave is launched into it.  
Our direct-lattice-transformation design follows four guidelines, (i)-(iv):

\noindent(i) First, we choose a mass density and adapt the lattice correspondingly. We then place it in a homogeneous isotropic medium surrounded by perfectly matched layers (implemented using the procedure described in Ref.\cite{Diatta2014}) and investigate the scattering of the object. 
We stress that it is not simple to launch a wave and extract the scattered field in mechanics. Clearly, one cannot simply pre-describe the time-dependent displacement-vector field at some boundary because this would fix the total field, i.e., incident field plus scattered field, rather than the incident field. In order to minimize the scattered field modification by the reflected wave we thus decide to use a line force excitation.\\
\noindent(ii) Second, we create an obstacle case for comparison. We choose a stress-free hole cut out in the homogeneous honeycomb lattice as the obstacle. \\
\noindent (iii) Third, for a cloak with outer radius $r_2$, the mass contained within this radius must be the same as for the reference lattice within the same radius. \\
\noindent (iv) The phase velocity in presence of the cloak must be the same as the phase velocity of the reference lattice. To meet this criterion, we consider the band structure of a periodic arrangement of cloaks (see Figure 2). We set all parameters $w$ and $W$ inside of the cloak to be the same and adjust $w$ and $W$ to meet the conditions (iii) and (iv).\\

\section{Numerical analysis of cloaking efficiency}

Once the four guidelines (i)-(iv) have been fulfilled, we compare the resulting behavior of cloak, obstacle, and reference under time-harmonic excitation by a line source at angular frequency $\omega$ (see Figure 2). Importantly all elements, i.e. homogeneous media and lattices are meshed and computed using finite elements (COMSOL PDE moduls with MUMPS solver).
In absence of a source, the displacement-vector field  ${\bf u}$ satisfies 
\begin{eqnarray}\label{nav1}
\nabla\cdot \left[ {\mathbf C} :\nabla  {\bf  u}
 \right]  + \rho\omega^2{\bf  u}={\bf 0}
%\label{nav1}
\end{eqnarray} 
with the mass density $\rho$ and the 4-order elasticity tensor ${\mathbf C}=(C_{ijkl})$, which satisfies Hookes' law $C_{ijkl}=\lambda\delta_{ij}\delta_{kl}+\mu(\delta_{ik}\delta_{jl}+\delta_{il}\delta_{jk})$, $i,j,k,l=1,2,3$, with $\lambda$ and $\mu$ being the Lam\'e coefficients, and where $\delta_{ij}=1$ if $i=j$ and 0 otherwise, is the Kronecker symbol. In our computations, we obviously assume continuity of displacement field and normal stress across all interfaces (this is automatically satisfied as (1) is solved in weak form in COMSOL). Furthermore, we consider $\mu=1$~GPa (the shear modulus), $\lambda=4.3$ GPa (the first Lamé modulus), and the mass density $\rho =10^3$ kg.m$^ {-3}$ for the constituent material of the lattice structure. The effective medium is described by $\mu_0=4.8$ MPa, $\lambda_0=180$ MPa, and a mass density $\rho_0 =1.98 \times 10^2$ kg.m$^ {-3}$.

\begin{figure}[h!]
\centering
	\includegraphics[width=11cm,angle=0]{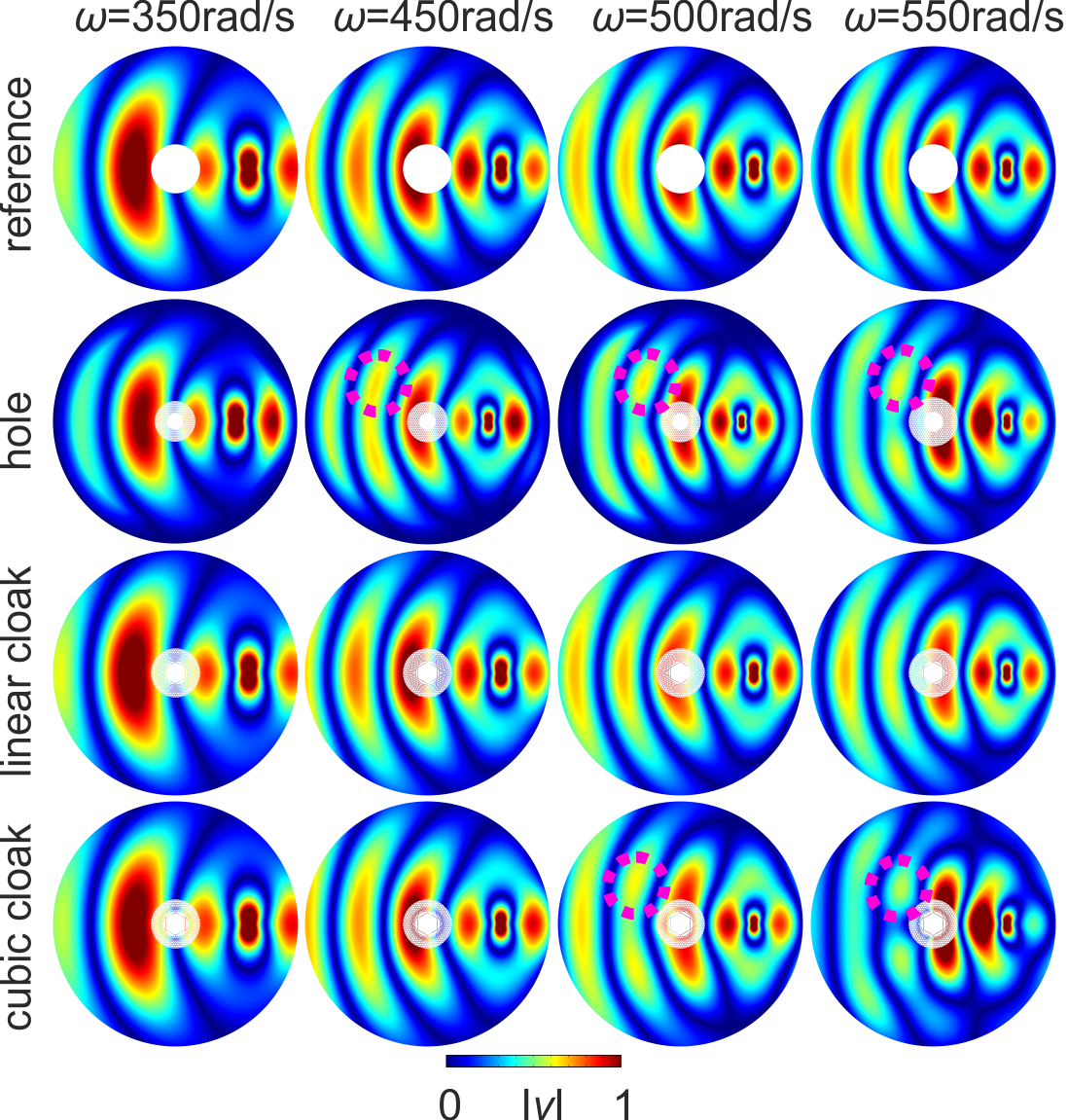}
	\caption{Displacement field map for different frequencies. Here, $\rm v$ is the y-component. The dotted circles are guides for the eyes.}
\label{Figure3}
\end{figure}

\begin{figure}[h!]
	\centering
	\includegraphics[width=8cm,angle=0]{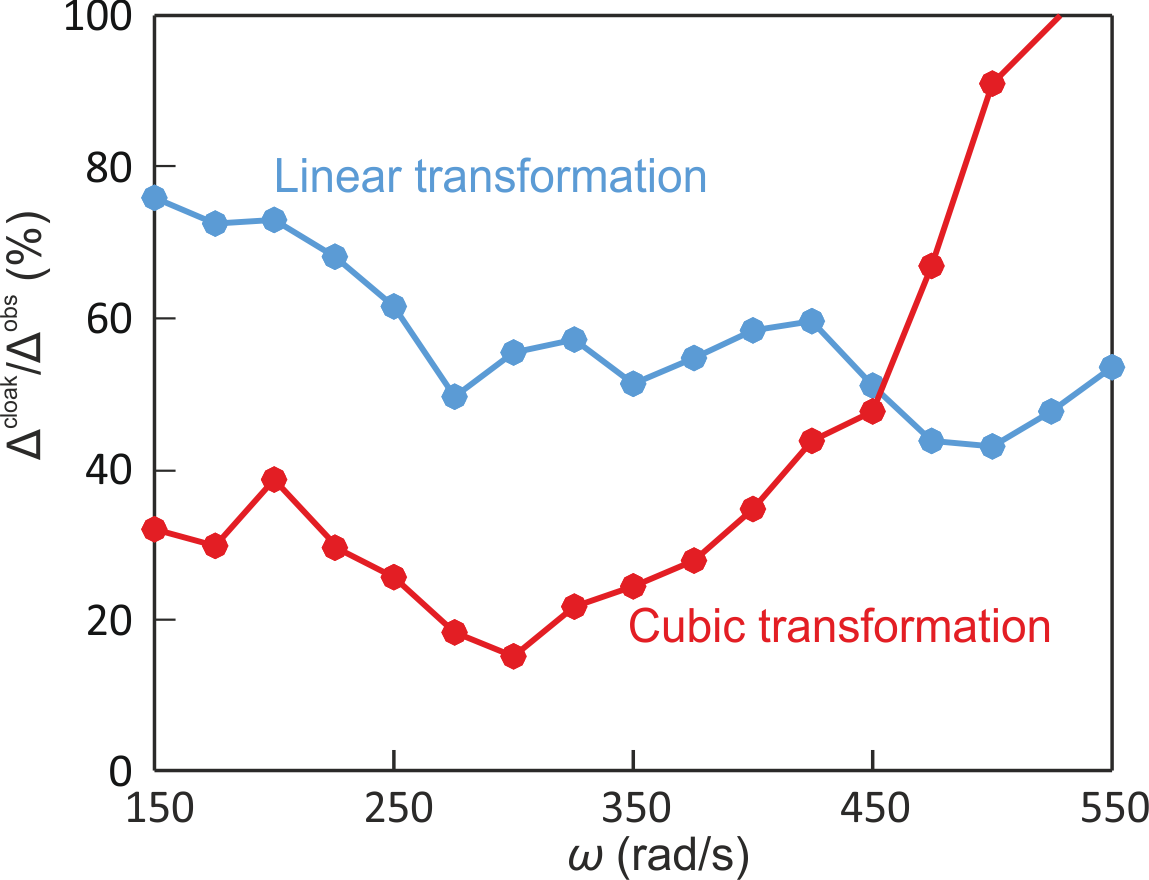}
	\caption{Focus on cloaking efficiency versus angular frequency. Scattered-like field (obtained by substracting the field of homogeneous sample to the cloaking and the holey sample) is plotted. $\Delta^i=\sqrt{\sum ({\bf u}_i-{\bf u}_r)^2}/\sqrt{\sum ({\bf u}_r)^2}$ with the index $i$ which corresponds to the obstacle (obs) or the cloak (cloak), and $r$ stands for the reference.}
	\label{Figure6}
\end{figure}

Let us start our discussion by inspecting the dispersion properties of the lattice cloak. To do so, we assume a doubly periodic set of lattice cloaks, in which case we encapsulate the lattice cloak inside a periodic cell with Floquet-Bloch periodic boundary conditions. The shear (green) and compression (red) acoustic bands are linear up to angular frequency $\omega=400$ rad/s, where the shear acoustic band meets first edge of the Brillouin zone, and thus undergoes some band folding. Below this frequency, one can consider the structure as an effective medium. On the other hand, for compressional waves, the effective medium-approximation is valid up to about $\omega=600$ rad/s, where a standing wave occurs (flat band) due to the inner resonance of the lattice cloak (also see localized eigenmode in (3) and flat dispersion curve in Figure \ref{Figure2}). Above $\omega=500$ rad/s (i.e., above the shaded gray area), a number of localized modes appear suggesting poor cloaking efficiency. Surprisingly, localized eigenmodes (see dispersion curve in Figure \ref{Figure2}) also occur in the effective-medium regime of the compressional waves. This unexpected behavior prevents cloaking for pressure waves in the long-wavelength limit. 

In contrast, dynamic cloaking for in-plane-shear waves is achieved with an experimentally accessible design. 
Figure 3 shows the performance of the four structures, namely the reference, the obstacle, and the linear and cubic cloaks under line-source excitation. At a low frequency of 350\,Hz, the forward scattered field is compensated in both linear and cubic cloaks and the behavior resembles that of the reference. As the frequency is increased to 450 and 550\,Hz, the cloaks still lead to reduced scattering.

To quantify the behavior, we plot in Figure 4 the mean standard deviation (over the computational domain) between cloak and reference, divided by the mean standard deviation between obstacle and hole. It becomes apparent that the cloak following the cubic coordinate transformation performs better than the one derived from the linear coordinate transformation. In the best case, scattering is reduced by 80\%, i.e., by a factor of five for the cubic cloak.

\section{Conclusion}
In summary, we have shown that the direct-lattice-transformation design approach applied to elastic micro-lattices (or metamaterials) can be expanded from the static to the dynamic case -- at least for in-plane-shear waves. The challenge of future experiments will lie in performing clean wave experiments on such structures. After all, obtaining perfectly matched layers at the boundaries of a sample based on non-symmetric Cosserat-material tensor distributions was even a major hurdle for the numerical calculations presented in this work. 

\section*{Acknowledgements}
We acknowledge Richard Craster (Imperial College), Alexander Movchan (University of Liverpool), Daniel Colquitt (University of Liverpool), Vincent Pagneux (LAUM) and Vincent Laude (CNRS) for discussion. This project has been performed in cooperation with the EIPHI Graduate School (contract ''ANR-17-EURE-0002'') and this work was also supported by the French Investissements d'Avenir program, project ISITE-BFC (contract ANR-15-IDEX-03). M.W. acknowledges support by the VIRTMAT project within the Helmholtz program "Science and Technology of Nanosystems" (STN) and by the Excellence Cluster "3D Matter Made to Order".
% and the Karlsruhe Institute of Technology VIRTMAT program.

%\section*{References}

\end{document}